\documentclass[runningheads]{svmult}

\usepackage{makeidx}   
\usepackage{graphicx}  
\usepackage{subeqnar}  
\usepackage{multicol}  
\usepackage{physprbb}  
\makeindex             

\begin{document}
\title*{Echoes in X-ray binaries; mapping the accretion flow}
\toctitle{Echoes in X-ray binaries; mapping the accretion flow}
%
%
\titlerunning{Echoes in X-ray binaries; mapping the accretion flow}
%
\author{Kieran O'Brien and Keith Horne}
\authorrunning{Kieran O'Brien and Keith Horne}
%
%
\institute{University of St Andrews, St Andrews, UK KY16 9SS}
%
%
\newcommand{\novasco}{GRO\,J1655-40} 
\newcommand{\novaper}{GRO\,J0422+32}
\newcommand{\novamus}{X-ray Nova Muscae 1991} 
\newcommand{\sco}{Scorpius~X-1} 
\newcommand{\Her}{Hercules~X-1}
\newcommand{\her}{Hercules~X-1}
\newcommand{\cyg}{Cygnus~X-2}
%
%
\newcommand{\HST} {\textit{HST}}
\newcommand{\xte} {\textit{RXTE}} 
\newcommand{\rxte}{\textit{RXTE}}
\newcommand{\XTE} {\textit{RXTE}} 
\newcommand{\RXTE}{\textit{RXTE}}
\newcommand{\GRO} {\textit{GRO}} 
%
%
\newcommand{\HI} {H\,\textsc{i}} 
\newcommand{\HII} {H\,\textsc{ii}} 
\newcommand{\HeI} {He\,\textsc{i}} 
\newcommand{\HeII} {He\,\textsc{ii}}
\newcommand{\HeIII}{He\,\textsc{iii}} 
%
%
\newcommand{\EBV}{E(B-V)} 
\newcommand{\Rv} {R_{\rm V}} 
\newcommand{\Av} {A_{\rm V}} 
%
%
\newcommand{\lam} {$\lambda$}
\newcommand{\lamlam}{$\lambda\lambda$} 
%
%
\newcommand{\comm}[1]{\textit{[#1]}} 
\newcommand{\etal}{{\it et al}.\ } 
\newcommand{\mdot}{\dot{M}}
\newcommand{\msolar}{M_{\odot}}
\newcommand{\degree}{$^{\circ}$}

\maketitle              

\begin{abstract}
In X-ray binaries much of the optical/UV emission arises from X-rays
reprocessed by material in the accretion disk, stream and the companion
star. The resulting reprocessed variability will be delayed in time with
respect to the X-ray variability by an amount depending on the position of
the
reprocessing regions. We can determine a range of time-delays present in
the
system. This time-delay transfer function can be used to `echo-map' the
geometry of the reprocessing regions in the binary system.

We present our modeling of this transfer function and show results from
our echo-mapping campaign using X-ray lightcurves from
RXTE, simultaneous with HST. In the X-ray transient, GRO j1655-40, shortly
after the 1996 outburst, we find evidence for reprocessing in the outer
regions of a thick accretion disk. 
\end{abstract}

\section{Introduction}
X-ray binaries (XRBs) are close binaries that contain a relatively
un-evolved donor star and a neutron star or black hole that is thought to
be accreting material through Roche-lobe overflow. Material passing
through the inner Lagrangian point moves along a ballistic trajectory
until impacting onto the outer regions of an accretion disk. This material
spirals through the disk, losing angular momentum, until it accretes onto
the central compact object, where X-rays are emitted from inner disk
regions. 

Much of the optical emission in XRBs arises from reprocessing of X-rays by
material in regions around the central compact object. Light travel times
within the system are of order 10s of seconds. Optical variability may
thus be delayed in time relative to the X-ray driving variability by an
amount characteristic of the position of the reprocessing region in the
binary, which depends in turn on the geometry of the binary. The optical
variability may be
modelled as a convolution of the X-ray variability with a
time-delay transfer function.

This time delay is the basis of an indirect imaging technique, known as
echo tomography, to probe the structure of accretion flows on scales that
cannot be imaged directly. 
Echo mapping has already been developed to interpret lightcurves of Active
Galactic Nuclei (AGN), where time delays are used to resolve photoionized
emission-line regions near the compact variable source of ionizing
radiation in the nucleus (Horne, this volume). In AGN the timescale of
detectable variations is
days to weeks, giving a resolution in the transfer functions of 1-10 light
days \cite{krolik91,kdh91}. In XRBs the binary separation is light seconds rather than light days, requiring high-speed
optical/UV
and X-ray lightcurves to probe the structure of the components of the
binary in detail. The
detectable X-ray and optical variations in the lightcurves of such systems
are also suitably fast. 

We present a simple geometric model for the time-delay
transfer functions of XRBs, using a synthetic binary code. We analyze
correlated X-ray and UV variability in \novasco, using our
computed transfer functions, to constrain the size, thickness and
geometric shape of the accretion disk in the system. 

\section{Reprocessing of X-rays} 
\label{reprocessing} 

In the standard model of reprocessing, X-rays are emitted by material in
the deep potential well of the compact object. These photoionize and heat
the surrounding regions of gas, which later recombine and cool, producing
lower energy photons. The optical emission seen by a distant observer is
delayed in time of arrival relative to the X-rays by two mechanisms. The
first is a finite reprocessing time for the X-ray photons, which
for this work is assumed to be
negligible \cite{obrien2000} and the second is the light travel times
between the X-ray source and the reprocessing sites within the binary
system. 

\subsection{light travel times}

The light travel times arise from the time of flight differences for
photons that are observed directly and those that are reprocessed and
re-emitted before travelling to the observer. These delays can be up to
twice the binary separation, obtained from Kepler's third law, 
\begin{equation}
\frac{a}{c} = 9.76\mbox{s}\left(\frac{M_{\rm{x}}+M_{d}}{\msolar}\right)^{\frac{1}{3}}  \left(\frac{P}{\mbox{days}} \right)^{\frac{2}{,3}}
\end{equation}
where $a$ is the binary separation, $M_{\rm{x}}$ and $M_{d}$ are the masses of
the compact object and donor star, $P$ is the orbital period. In LMXBs the
binary separation is of the order of several light seconds.

The time delay $\tau$ at binary phase $\phi$ for a reprocessing site with
cylindrical coordinates (R, $\theta$, Z) is
\begin{equation}
\label{delayeqn}
\tau(\vec{x},\phi) = \frac{\sqrt{R^2 + Z^2}}{c}(1+\sin{i}\cos(\phi-\theta)) -\frac{Z}{c}\cos{i}
\end{equation}
where $\it{i}$ is the inclination of the system and $\it{c}$ is the speed
of light. 

The dynamic response function is found by considering how a change in
X-ray flux, $\Delta\,f_{\rm{x}}(t)$, drives a change in the reprocessed
flux.
We can define the
dynamic time delay transfer function to be

\begin{eqnarray}
\Psi_{\nu}\left(\lambda,\tau,\phi\right) & = & \int \left[ \frac{\delta I_{\nu}\left(\lambda,\vec{x}, \Delta f_{\rm{x}}(t-\tau) \right)}{\delta f_{\rm{x}}(t-\tau)}\right] \, d \Omega(\vec{x},\phi) \, \delta (\tau - \tau(\vec{x},\phi))\,,
\label{dynamic}
\end{eqnarray} 
where $\tau(\vec{x},\phi)$ is the geometric time delay of a reprocessing
site at position $\vec{x}$, see (\ref{delayeqn}). 

\section{Model X-ray Binary code}
\label{code}

We have developed a code to model time delay transfer functions based on
determining the contributions from different regions in the binary. In
this section we describe the models used to construct the individual
regions of the binary; the donor star, the accretion stream and the
accretion disk. The code uses distances scaled to the binary separation in
a right-handed cartesian coordinate system corotating with the binary. 
Each surface panel is a triangle, characterized by its area
$dA$, orientation $\vec{n}$, position $\vec{x}$ and temperature $T$.

\subsection{Donor Star}
\label{star}

The donor star is modeled assuming it fills its critical Roche potential,
so that mass transfer occurs via Roche lobe overflow through the inner
Lagrangian point. Optically thick panels are placed over the surface of
the Roche potential. The panels are triangular so that the curved surfaces
of the binary are mapped more accurately than is possible using 4-sided
shapes \cite{rutten94}. The panels, when unirradiated,
are assigned an effective temperature $T_{star}$ given by the spectral
type of the donor star.

\subsection{Accretion stream}
\label{stream}

The accretion stream is modeled by following the ballistic trajectories of
4 test particles.
The `width' and the `height' of
the stream (its extent in the y- and z-directions respectively) is determined by the initial
positions of the test particles. The stream is symmetric about the x-y plane.
The unirradiated accretion stream is assumed to have a constant
temperature $T_{s}$ along it's length and the effects of irradiation are
considered in the same way as those of the donor star. 
\subsection{Accretion disk}
\label{disk}
The disk thickness is assumed to increase with radius from 0 at
$\mbox{R}=\mbox{R}_{in}$ to H$_{out}$ at $\mbox{R}=\mbox{R}_{out}$, with the
form,
\begin{equation}
H= R_{out} \left( \frac{H}{R}\right )_{out}\left( \frac{R-R_{in}}{R_{out}-R_{in}}\right )^{\beta},
\end{equation}
where the parameters are the inner and outer disk radii, $R_{in}$ and
$R_{out}$ in units of R(L1), the half thickness of the outer disk
$(H/R)_{out}$ and the exponent $\beta$ which describes the overall shape
of the disk. The temperature structure of the un-irradiated disk is that
of a steady state disk, in the absence of irradiation,
\begin{equation}
\label{disktemp}
T_{disk}(R)=T_{out}\left( \frac{R}{R_{out}}\right)^{-\frac{3}{4}},
\end{equation}
where $T_{out}$ and $T_{in}$ are the temperatures of the outer and inner
disk respectively. 

\subsection{Irradiation model}
\label{irrad}
The effective temperature of a region at a distance R from the X-ray
source, assumed in our model to be a point source located at the centre of
the accretion disk, is found from the accretion luminosity for a typical
LMXB,
\begin{equation}
\label{tx}
T_{\rm{x}}^{4} = \frac{L_{\rm{x}} (1-A)}{4\pi \sigma R^2}
\end{equation}
and 
\begin{equation}
L_{\rm{x}} = \eta \frac{ GM_{\rm{x}}\mdot}{R_{ns}}\,,
\end{equation}
where $T_{\rm{x}}$ is the temperature, $A$ the albedo, $\eta$ the
efficiency, $M_{\rm{x}}$, the mass of the compact object, $\mdot$ the accretion
rate onto the compact object, $R_{ns}$ is the size of the compact object
and $R$ is the distance between the compact object and the irradiated
panel. This is normalised using the binary separation, a, the distance
between the centres of mass of the stars, as is the coordinate system for
the binary. 

The irradiation of the binary takes place in three stages. The first stage
is to calculate the temperature structure of the binary in the absence of
any irradiation. This is done with characteristic temperatures for the
donor star and the accretion stream and disk. 
The temperature structure of the disk is assumed to that for an
unirradiated disk as given in~(\ref{disktemp}). The surface
panels of the binary exposed to X-rays are determined by projecting the
binary surfaces onto a spherical polar representation of the sky, as it
appears from the X-ray source. Each triangular panel is mapped to the
sky starting with the one furthest from the source and ending with the
panel closest. Those panels remaining visible and unocculted on the
sky map are irradiated. The change in effective temperature of a panel
is scaled by the projected area with respect to the X-ray source at a
distance $R$ from the source. Hence the temperature after irradiation is
given by,

\begin{equation}
\label{tempx}
T^{4} = T_{\rm{x}}^{4} \cos\theta_{\rm{x}} \left( \frac{a}{R} \right)^{2} + T_{eff}^{4}
\end{equation}
where $T$ is the temperature of the panel, $\theta_{\rm{x}}$ the angle between
the line of sight from the central source and the normal to the surface of
the panel and $T_{eff}$ is the unirradiated effective temperature of the
panel. 

\begin{figure}[ht]
\begin{center}
\includegraphics[width=.8\textwidth,angle=0.]{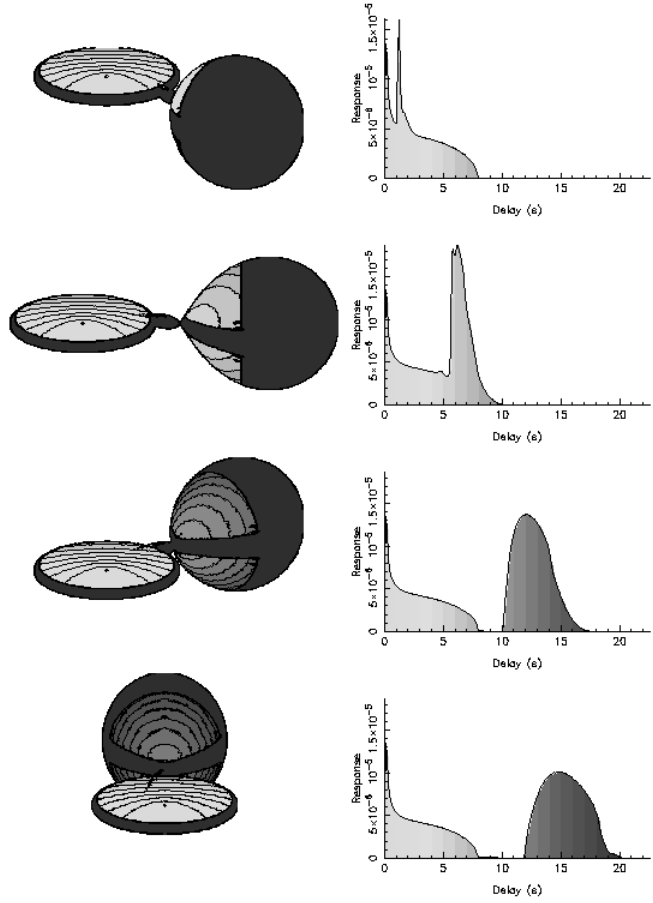}
\end{center}
\caption[]{Left, model X-ray binaries, based on the Scorpius~X-1 binary parameters, showing iso-delay surfaces projected onto the irradiated surfaces of the binary. Right, the associated time delay transfer functions, showing the relative contributions from the regions highlighted in the model X-ray binaries.}
\label{binaries}
\end{figure}

The second stage is to irradiate the binary with the constant component of
the X-ray flux. This component of the X-ray flux is equated to the mean
effective temperature of the X-ray source, as given in~(\ref{tempx}), where $T_{\rm{x}} \equiv \overline{T_{\rm{x}}}$. The third and
final stage is to repeat stage two with $T_{\rm{x}} \equiv {T(t)_{\rm{x}}}$, which
represents irradiating the binary with a time varying component. The
difference between stages two and three represents the temperature change
of the panels due to the time varying component of the X-ray flux alone,
$\Delta f_{\nu}(t)$. 

The irradiated regions of the binary can be clearly seen in Fig.~\ref{binaries}, where the left hand panels show a typical X-ray binary, using binary parameters based on those of \sco, viewed from an inclination of 60\degree.

The response of a panel to the variable component of the irradiating
X-ray flux is given by,
\begin{eqnarray}
I_{\nu}\!\left(\lambda,\vec{x},\Delta f_{\rm{x}}(t)\right) & = & \!\int\! \left[ B_{\nu}\!\left(\lambda,T_{\rm{x}}(t)\right) - B_{\nu}\!\left( \lambda, \overline{T_{\rm{x}} } \right) \right] P\!\left(\lambda\right) I\!\left(u,\alpha\right) d\Omega\!\left(
 \vec{x} , \phi\right) d\lambda.
\end{eqnarray}

\begin{figure}[ht]
\begin{center}
\includegraphics[height=.8\textwidth,angle=270.]{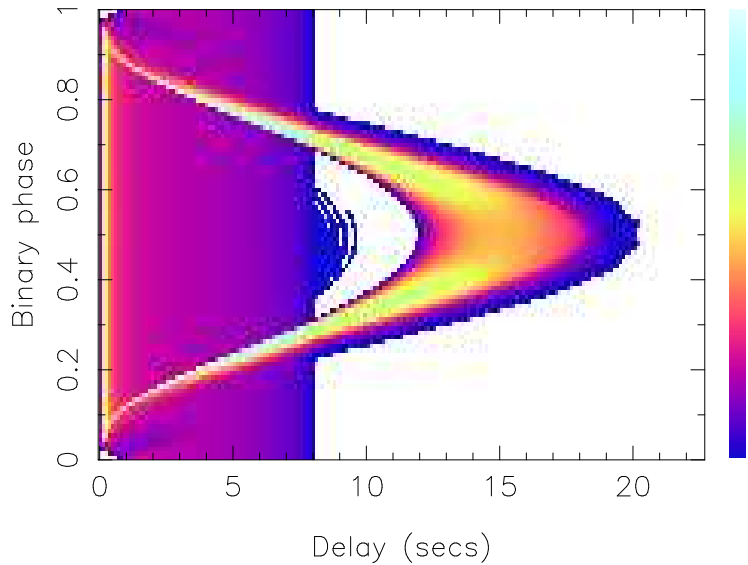}
\end{center}
\caption[]{A plot of time-delay transfer functions as a function of binary
phase, based on the binary parameters of Scorpius~X-1 \cite{vrtilek91,kallman91}. The accretion disk has constant time delays in the
region 0-8 seconds, whereas the time delays from the companion star are
seen to vary sinusoidally with binary phase between 0-20 seconds.}
\label{echophase}
\end{figure}

This response is substituted into the expression for the dynamic response
given in~(\ref{dynamic}). The resulting time delays are mapped onto a time
delay grid to produce our time-delay transfer function. Examples of this
transfer function can be seen in the right-hand panels of
Fig.~\ref{binaries}. The time delay of donor star can be seen to change
with binary phase in the individual images, whereas the accretion disk
remains constant throughout the binary orbit. This effect can be seen more
clearly in Fig.~\ref{echophase}, where the individual transfer functions
have been trailed in orbital phase to create an echo-phase diagram,
showing clearly how
the time delays of the individual components change with binary phase. 

\section{Results for \novasco}

\begin{figure}[ht]
\begin{center}
\includegraphics[height=.8\textwidth,angle=270.]{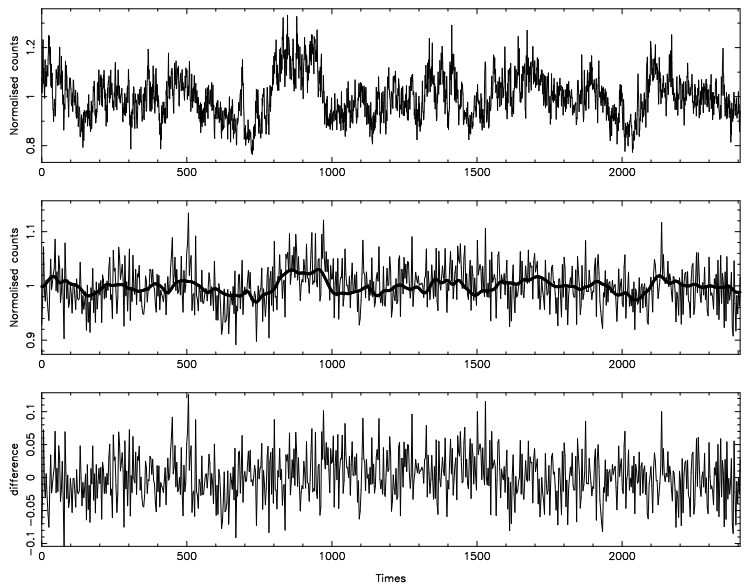}
\end{center}
\caption[]{The best-fit results for GRO J1655-40 from a grid-search of trial transfer functions. Top panel, the normalised X-ray driving lightcurve
from RXTE. Middle panel, UV lightcurve from HST with synthetic UV
lightcurve superimposed (thick line). Bottom panel, residuals of the fit
to the UV lightcurve.}
\label{lightcurves}
\end{figure}

\begin{figure}[ht]
\begin{center}
\includegraphics[height=.8\textwidth,angle=270.]{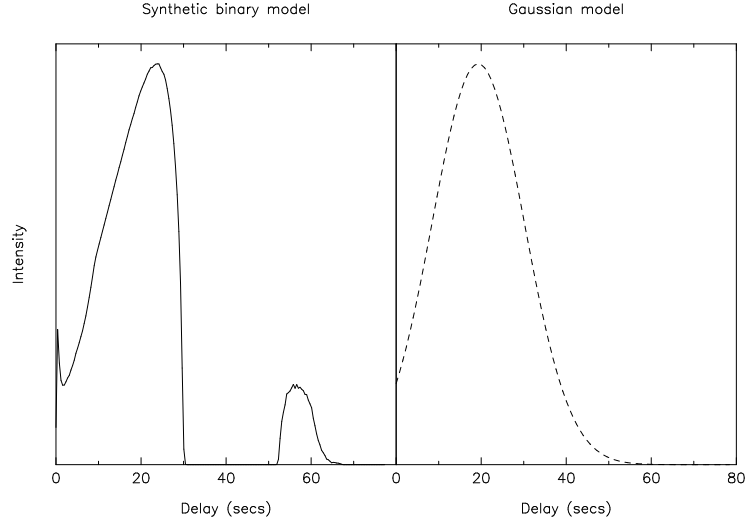}
\end{center}
\caption[]{A comparison of the best fit transfer functions for GRO~J1655-40
from our two modeling methods. On the left is the synthetic X-ray binary
model transfer function and on the right is the acausal Gaussian transfer
function.}
\label{trans}
\end{figure}

We have used these model transfer functions to constrain the geometric
parameters for the soft X-ray transient \novasco. The X-ray data was taken
with the PCA onboard \RXTE\ on June 8 1996, simultaneous with the \HST\
data. These lightcurves are shown in the top and middle panels of
Fig.\ref{lightcurves} respectively. The data were previously fitted with
causal and
acausal Gaussian transfer functions and found to have a mean 14.6 $\pm$
1.4 seconds, with a RMS of 10.5 $\pm$ 1.9 seconds \cite{hynes98echo}. We
have used our model X-ray binary code to predict transfer functions,
using the known binary parameters for \novasco\,\cite{orosz97}. The trial
transfer functions are convolved with the X-ray lightcurve to create a
synthetic reprocessed lightcurve. The badness of
 fit is found between the synthetic and observed reprocessed lightcurves and minimized to find the best-fit values for the size, thickness and shape of the accretion disk. The synthetic lightcurve from the best-fit transfer function is shown by the thick 
line in the middle panel of Fig.~\ref{lightcurves}, while the bottom panel shows the residuals of the fit. The best-fit model transfer function is shown in Fig.~\ref{trans}, together with the best-fit Gaussian transfer function from \cite{hynes98echo}.

We find that the disk extends to 67\% of the way to the inner Lagrangian point, is geometrically thick, with an opening angle $\sim$ 14\degree and is somewhat flared. This has the effect of shadowing the donor star from much of the irradiation. 

\section{Discussion}

We have used the time delays observed between the X-ray and optical/UV
variability in X-ray binaries to echo-map the irradiated regions. We have
developed a code to simulate the time-delay transfer functions for such
systems and find that, in the case of the SXT, \novasco, there is evidence
for a geometrically thick outer accretion disk that shields the innerface
of the donor star from irradiation.

While the method of echo-tomography of X-ray binaries is still in its
infancy, we have shown that with just a small amount of data, from
co-ordinated observing campaigns using ground-based and satellite
observatories, this technique can reveal interesting insights into the
geometry of X-ray binaries. Furthermore this technique has the promise of
probing the structure and geometry of such systems on scales unobtainable
with any other current technique.

%

\end{document}